\documentclass[aps,prc,twocolumn,fleqn,floatfix,superscriptaddress, nofootinbib]{revtex4}

\usepackage{epsfig}
\usepackage[usenames]{color} 
\usepackage{amsmath,amssymb,amsfonts}

\newcommand \etc {{\it etc.} }
\newcommand \tie {{\it i.e.}}

\newcommand \kd {\delta}
\newcommand \ra {\rightarrow}

\newcommand \w {\omega}

\newcommand \vk {\vec{k}}

\newcommand \vx {\vec{x}}
\newcommand \vp {\vec{p}}
\newcommand \vecr {\vec{r}}

\newcommand \g {\gamma}

\newcommand \ep {\epsilon}

\newcommand \p {^{\prime}}

\newcommand \x {\cdot}
\newcommand \hf {\frac{1}{2}}
\newcommand \A {\alpha}

\newcommand \lc {\langle}
\newcommand \rc {\rangle}
\newcommand \prt {\partial}
\newcommand \D {\Delta}
\newcommand \sg {\sigma}

\newcommand \nt {\noindent}

\newcommand {\llb} { \left[ \frac{\mbox{}}{\mbox{}} \right.}
\newcommand {\lrb} { \left. \frac{\mbox{}}{\mbox{}} \right] }

\newcommand \bvec{\left( \begin{array}{c} }
\newcommand \evec{\end{array} \right)}

\newcommand \eg {{\it e.g.}}
\newcommand \bea{\begin{eqnarray} }
\newcommand \eea{\end{eqnarray} }
\newcommand \nn {\nonumber}
\newcommand {\be} {\begin{equation}}
\newcommand {\ee} {\end{equation}}

\newcommand {\mbx} {\mbox{}}

\newcommand{\ata} {&\times&}
\newcommand{\psibar} {\bar{\psi}}

\newcommand{\qhat} {$\hat{q}$ }

\begin{document}

\title{Systematic Comparison of Jet Energy-Loss Schemes in a Realistic Hydrodynamic Medium}

\author{Steffen A. Bass}
\affiliation{Department of Physics, Duke University, Durham, NC 27708, USA}

\author{Charles Gale}
\affiliation{Department of Physics, McGill University, H3A 2T8, Montreal, Quebec, Canada}

\author{Abhijit Majumder}
\affiliation{Department of Physics, Duke University, Durham, NC 27708, USA}

\author{Chiho Nonaka}
\affiliation{Department of Physics, Nagoya University, Nagoya 464-8602, Japan}

\author{Guang-You Qin}
\affiliation{Department of Physics, McGill University, H3A 2T8, Montreal, Quebec, Canada}

\author{Thorsten Renk}
\affiliation{Department of Physics, PO Box 35 FIN-40014 University of Jyv\"{a}skyl\"{a}, Finland}
\affiliation{Helsinki Institute of Physics, PO Box 64 FIN-00014 University of Helsinki, Finland}

\author{J\"org Ruppert}
\affiliation{Department of Physics, McGill University, H3A 2T8, Montreal, Quebec, Canada}

\date{\today}

\begin{abstract}
We perform a systematic comparison of three different jet energy-loss approaches. These include the 
Armesto-Salgado-Wiedemann scheme based on the approach of Baier-Dokshitzer-Mueller-Peigne-Schiff and Zakharov (BDMPS-Z/ASW), 
the Higher Twist approach (HT) and a scheme based on the approach of Arnold-Moore-Yaffe (AMY). In this comparison, 
an identical medium evolution will be utilized for all three approaches: not only does this entail the use of the same realistic 
three-dimensional relativistic fluid dynamics (RFD)  simulation, but also includes the use of identical initial parton-distribution functions and final fragmentation functions. We are, thus, in a unique position, not only to isolate
fundamental differences between the various approaches, but also to make rigorous calculations for different 
experimental measurements using ``state of the art'' components.
All three approaches are reduced to a version which contains 
only one free tunable parameter, this is then related to the well known transport parameter $\hat{q}$.
We find that the parameters of all three calculations can be adjusted to provide a good
description of inclusive data on $R_{AA}$ versus transverse momentum. However, we do observe slight differences in
their predictions for the centrality and azimuthal angular dependence of $R_{AA}$ vs. $p_T$. We also note that the
value of the transport coefficient $\hat{q}$ in the three approaches to describe the data differs significantly. 
\end{abstract}
\maketitle

\section{Introduction}
The first seven years of operations at the Relativistic Heavy-Ion
Collider (RHIC), performing collisions of gold nuclei
at  $\sqrt{s}_{NN}=130$~GeV and $\sqrt{s}_{NN}=200$~GeV, 
have yielded a vast amount of interesting and
sometimes surprising results \cite{Adcox:2004mh,Back:2004je,Adams:2005dq,Arsene:2004fa}. 
Many of these have not yet been fully evaluated or understood by theory.
There exists mounting evidence that RHIC has created
a hot and dense state of deconfined QCD matter with properties similar to 
that of an ideal fluid \cite{Gyulassy:2004zy} -- this state of matter 
has been termed the {\em strongly interacting Quark-Gluon-Plasma} (sQGP).

RHIC has generated a wealth of experimental data on high momentum 
hadron emission, including, but not limited to, the nuclear modification 
factor $R_{AA}$, its modification as a function of the reaction plane 
(a measure of the 
azimuthal anisotropy of the cross section) and a whole array of high-$p_T$ 
hadron-hadron correlations. In these observables, one compares the ratio of certain yields in a 
heavy-ion collision to those in a  $p$-$p$ collision, either scaled up by the number of 
expected binary collisions, e.g., for the single hadron
suppression factor $R_{AA}$, or directly, as in the case of triggered distributions of associated hadrons, e.g., the $I_{AA}$ \cite{Adler:2003kg,Adler:2003cb,Adams:2003am}. 
Experimental data for most of these observables exist as functions of rapidity 
and centrality, for a wide range of $p_T$ of the produced particle or particles.

The emission of hadrons with large transverse momentum is observed to
be strongly suppressed in central collisions of heavy nuclei \cite{Adcox:2001jp,Adler:2002xw}. 
The origin of this phenomenon, commonly referred to as {\em jet-quenching}, can
be understood in the following way:
during the early pre-equilibrium stage of the relativistic heavy-ion
collision, scattering of partons which leads to the formation
of deconfined quark-gluon matter often engenders large momentum transfers 
which leads to the formation of two back-to-back hard partons.  These
traverse the dense medium, losing energy and finally
fragment into hadrons which are observed by the experiments.
Within the framework of perturbative QCD, the process with largest energy loss of a fast parton is gluon radiation
induced by collisions with the quasi-thermal medium~\cite{Wang:1991xy,Gyulassy:1993hr,Wang:1994fx,Baier:1996kr,Baier:1996sk,Baier:1998yf,Zakharov:1996fv,Zakharov:1997uu,Zakharov:1998sv}.

Computations of jet modification have acquired a
certain sophistication as regards the
incorporation of the partonic processes involved.
However, the role of the medium has often been
relegated to the furnishing of an overall density and its
variation with time~\cite{Gyulassy:2001nm,Salgado:2003gb,Jeon:2003gi,Wang:2003mm}. 
Notable departures from these
simple treatments include attempts  to  incorporate radial expansion, both
schematically~\cite{Majumder:2006we,Gyulassy:2001kr} as well as within a fireball evolution model~\cite{Renk:2005ta}. 
The first attempt to incorporate energy loss in a three dimensional (3-D) relativistic fluid dynamical (RFD) simulation was 
carried out by Hirano and Nara in Ref.~\cite{Hirano:2002sc}. In this effort, while a full 3-D RFD simulation was used, the 
energy loss of hard jets was carried out rather schematically. This approach was also extended to the case of two particle 
correlations in Ref.~\cite{Hirano:2003hq}.  In a later effort the authors also incorporated a simplified version of the 
Gyulassy-Levai-Vitev (GLV) energy loss formalism at leading order in opacity~\cite{Hirano:2003pw}.

Besides the 
simplified version of the GLV formalism used, the authors attempted to apply the results to the region in $p_T \leq 6$ GeV, which is
the region where data were available at the time. 
In spite of the success of Ref.~\cite{Hirano:2003pw} in explaining the suppression of single inclusive pions, 
such a formalism cannot address the flavor dependence of the elliptic flow in this region of $p_T$. It has since been established 
that jet fragmentation in vacuum is not the primary mechanism of hadronization in the range of $p_T < 6$ GeV and there is a sizeable component 
which arises from recombination. Current rigorous implementations of jet modification in dense matter require that the $p_T$ of the 
detected hadron be above 6 GeV. This allows for a treatment where the final hadronization may be treated using the standard 
vacuum fragmentation functions and the ability of a given energy loss formalism to compare with experimental data is dependent 
solely on the details of the interaction of the parton with the medium in that formalism. This allows for a comparison between 
formalisms where all other components of the calculation such as the initial parton distribution, the final fragmentation function as well as the 
space-time profile of the medium are identical. This article presents the first attempt to perform such a comparison between the 
remaining three formalisms: the BDMPS/ASW, the HT and the AMY approach.

Besides just a comparison between formalisms, this paper will simultaneously 
also apply the different formalisms in comparison to data. A realistic 
comparison with data requires a sophisticated model of the medium. 
The availability of a three-dimensional hydrodynamic evolution
code \cite{Nonaka:2006yn} allows
for a much more detailed study of jet interactions in a
longitudinally and transversely expanding medium.
The variation of the gluon density in such a medium
is quite different from  that in a simple Bjorken expansion.
This allows for a step-by-step approach to the study of jet-medium
interactions.
Over the past year we have already utilized our 
evolution model to provide the time-evolution of the medium produced at RHIC 
for jet energy-loss calculations performed in the BDMPS/ASW~\cite{Renk:2006sx}, 
HT~\cite{Majumder:2007ae} and AMY~\cite{Qin:2007zz} approaches.  
In each of the three projects, the inclusive as well as the azimuthally differential
nuclear suppression factor $R_{AA}$ of pions was studied as a function of their 
transverse momentum $p_T$. In addition, the influence of collective flow, variations 
in rapidity, and energy-loss in the hadronic phase were addressed for the selected 
approaches. 

In this manuscript, we shall perform a systematic comparison of jet energy-loss calculations in the
BDMPS/ASW, HT and AMY approaches. Since we use the same medium evolution in all three approaches we are in a position to isolate
differences among the three calculations solely due to their energy-loss schemes. 
This will allow us to answer the question whether the observed differences between the
different schemes (when compared to data) are due to differing treatment of the medium evolution
and its coupling to the energy-loss calculation or whether they are rooted in more fundamental issues related to the energy-loss schemes themselves, e.g. due to the approximations and
assumptions made when deriving the respective schemes.
In Sec.~II, we briefly review the 3D
hydrodynamical description of the medium. We then discuss in Sec.~III the theoretical setup of different energy loss
schemes and their connection to the 3D dynamical evolving medium. Numerical results are presented in comparison to the
RHIC data where already available in Sec.~IV. In Sec.~V, we discuss issues related to further comparisons of our calculations with 
the data on $R_{AA}$ versus the reaction plane and present concluding 
discussions and an outlook to future work in Sec.~VI.


\section{Hydrodynamic Description of the Medium}


Relativistic Fluid Dynamics (RFD, see e.g.
\cite{Bjorken:1982qr,Clare:1986qj,Dumitru:1998es})
is ideally suited for the high-density phase of heavy-ion reactions
at RHIC,
but breaks down in the later, dilute, stages of the
reaction when the mean free paths of the hadrons become
large and flavor degrees of freedom
are important. The biggest advantage  of  RFD
is that it directly incorporates an equation of state as input
and thus is so far the only dynamical model in which a phase
transition can
explicitly be incorporated. 
Starting point for a RFD calculation is the
relativistic hydrodynamic equation
\begin{equation}
\partial_\mu T^{\mu \nu} = 0,
\label{Eq-rhydro}
\end{equation}
where $T^{\mu \nu}$ is the energy momentum tensor which is given by
\begin{equation}
T^{\mu \nu}=(\epsilon + p) U^{\mu} U^{\nu} - p g^{\mu \nu}.
\end{equation}
Here $\epsilon$, $p$, $U$ and $g^{\mu \nu}$ are energy density,
pressure, four velocity and metric tensor, respectively.
The relativistic hydrodynamic equation Eq.\ (\ref{Eq-rhydro})
is solved numerically using baryon number $n_B$ conservation
\begin{equation}
\partial_\mu (n_B (T,\mu) U^\mu)=0.
\end{equation}
as a constraint and closing the resulting set of partial
differential equations by specifying an equation of state (EoS):
$\epsilon = \epsilon(p)$.
In the ideal fluid approximation
(i.e. neglecting off-equilibrium effects) and once the initial conditions
for the calculation have been fixed,
the EoS is the {\em only}
input to the equations of motion and relates directly to
properties
of the matter under consideration. Ideally, either the initial conditions or the
EoS should be determined beforehand by an ab-initio calculation (e.g. for the EoS via
a lattice-gauge calculation), in which case a fit to the data would allow
for the determination of the remaining quantity.
Our particular RFD implementation
utilizes a Lagrangian mesh and 
light-cone coordinates, i.e.,  $(\tau,x,y,\eta)$ where $\tau=\sqrt{t^2-z^2}$ is the proper time and $\eta$ is the pseudo-rapidity.
This is done in order to optimize the model for the ultra-relativistic regime
of heavy collisions at RHIC.

We assume that hydrodynamic expansion starts at
$\tau_0=0.6$ fm. Initial energy density and
baryon number density are parametrized by
\begin{eqnarray}
\epsilon(x,y,\eta)& =& \epsilon_{\rm max}W(x,y;b)H(\eta),
\nonumber \\
n_B(x,y,\eta)& = & n_{B{\rm max}}W(x,y;b)H(\eta),
\end{eqnarray}
where $b$ and  $\epsilon_{\rm max}$ ($n_{B{\rm max}}$) are
the impact parameter and the maximum value of energy density
(baryon number density), respectively.
$W(x,y;b)$ is given by a combination of wounded nuclear model and
binary collision model \cite{Kolb:2001qz} and  $H(\eta)$ is given
by $\displaystyle
H(\eta)=\exp \left [ - (|\eta|-\eta_0)^2/2 \sigma_\eta^2 \cdot
\theta ( |\eta| - \eta_0 ) \right ]$.
RFD has been very successful in describing single soft matter
properties at RHIC, especially
collective flow effects and particle spectra 
\cite{Kolb:2003dz,Huovinen:2003fa,Hirano:2002hv,Nonaka:2006yn}.
All parameters of our hydrodynamic evolution \cite{Nonaka:2006yn}
have been fixed
by a fit to the soft sector (elliptic flow, pseudo-rapidity
distributions and low-$p_T$ single particle spectra), therefore
providing us with a fully determined 
medium evolution for the hard probes to propagate through.


\section{Jet Energy-Loss Schemes}


The majority of current approaches to the energy loss of light partons may 
be divided into four major schemes often referred to by the names of the original 
authors:
\begin{itemize}
\item Higher Twist (HT)~\cite{Guo:2000nz,Wang:2001ifa,Zhang:2003yn,Majumder:2004pt,Majumder:2007hx,Majumder:2007ne}
\item Path integral approach to the opacity expansion (BDMPS-Z/ASW)~\cite{Zakharov:1996fv,Zakharov:1997uu,Zakharov:1998sv,Baier:1996kr,Baier:1996sk,Wiedemann:2000ez,Wiedemann:2000za,Salgado:2002cd,Salgado:2003gb,Armesto:2004ud}
\item Finite temperature field theory approach (AMY)~\cite{Arnold:2001ba,Arnold:2000dr,Jeon:2003gi,Turbide:2005fk}
\item Reaction Operator approach to the opacity expansion (GLV)~\cite{Gyulassy:1999zd,Gyulassy:2000er,Gyulassy:2001nm,Djordjevic:2003zk,Wicks:2005gt}
\end{itemize}
All schemes utilize a factorized approach where the 
final cross section to produce a hadron $h$  with transverse momentum $p_T$ 
(rapidity between $y$ and $y+dy$) 
may be expressed as a convolution of initial nuclear structure functions ($G_a^A(x_a),G_b^B(x_b) $, initial state 
nuclear effects such as shadowing and Cronin effect are understood to be included) to produce 
partons with momentum fractions $x_a,x_b$, a 
hard partonic cross section to produce a high transverse momentum parton $c$ with a 
transverse momentum $\hat{p}$ and a medium 
modified fragmentation function for the final hadron ($\tilde{D}_c^h(z)$), 

\bea
\frac{d^2 \sg^h}{dy d^2 p_T} &=& \frac{1}{\pi} \int dx_a \int d x_b G^A_a(x_a) G^B_b(x_b) \nn \\
\ata \frac{d \sg_{ab \ra cX} }{d \hat{t}} \frac{\tilde{D}_c^h(z)}{z}. \label{basic_cross}
\eea

\nt
In the vicinity of mid-rapidity, $z=p_T/\hat{p}$ and $\hat{t} = (\hat{p} - x_a P)^2$  ($P$ is the 
average incoming momentum of a nucleon in nucleus A).
The entire effect of energy loss is concentrated in the  calculation of the 
modification to the fragmentation function. The four models of energy loss are in a  
sense four schemes to estimate this quantity from perturbative QCD calculations.  
While the terminology (medium modification) used to describe the change in the fragmentation 
function seems to indicate that the medium has influenced the actual process of the formation of the final hadrons 
from the partonic cloud, 
this is not the case. All computations simply describe the change in the gluon radiation spectrum 
from a hard parton due to the presence of the medium. The final hadronization of the hard parton 
is always assumed to occur in the vacuum after the parton, with degraded energy, has escaped from the medium. 
Note that some of the hard gluons radiated from the hard parton will also encounter similar ``modification'' 
in the medium and may endure vacuum hadronization after escaping from the medium. Differences 
between formalisms also arise in the inclusion of hadrons from the fragmentation of such sub-leading gluons: 
whereas in approaches which compute the change in the distribution of final partons (such as AMY) or the change in the 
distribution of final hadrons (such as HT), hadrons from sub-leading gluons are implicitly included, formalisms which compute 
the energy loss of the leading parton  (such as ASW), do not include such sub-leading corrections.

\begin{figure}
\includegraphics[width=\columnwidth]{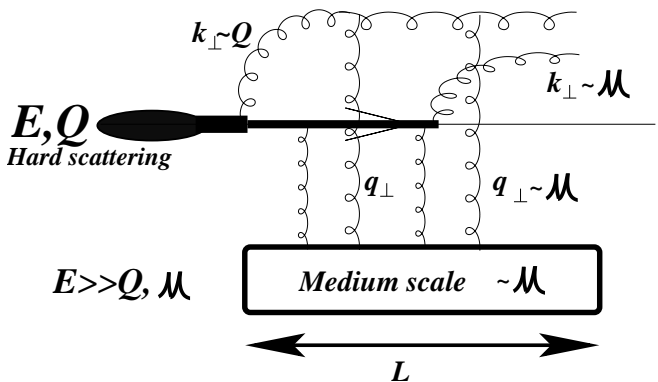}
\caption{A schematic picture of the various scales involved in the modification of jets in dense matter. }
\label{fig0}
\end{figure}

To better appreciate the approximation schemes, one may introduce a set of scales (see Fig.~\ref{fig0}):
$E$ or $p^+$, the forward energy of the jet, $Q^2$, the virtuality of the initial jet-parton, $\mu$, 
the momentum scale of the medium and $L$, its spatial extent. Most of the   
differences between the various schemes may be reduced to the different 
relations between these various scales assumed by each scheme as well 
as by how each scheme treats or approximates the structure of the medium. In 
all schemes, the forward energy of the jet far exceeds the medium scale, $E >> \mu$. The schemes 
are presented from one extreme of the approximation set (higher twist approach)  to the opposite extreme 
(finite temperature approach), similarities in intermediate steps of the calculation will not be repeated. In the following, we
shall focus on the first three listed approaches, for which
we shall present results in section~\ref{results} (note that
a calculation of GLV jet energy-loss in a 3D hydrodynamic medium
has been presented elsewhere \cite{Hirano:2003pw}).

\subsection{Higher Twist Formalism}


The origin of the higher twist (HT) approximation scheme lies in the calculations of 
medium enhanced higher twist corrections to the total cross section in  
Deep-Inelastic Scattering (DIS) off large nuclei~\cite{Qiu:1990xxa}. One re-sums 
power corrections to the leading twist cross sections, 
which, though suppressed by powers of the hard scale $Q^2$, are enhanced by the 
length of the medium. This technology of identifying and isolating power corrections 
is used to compute the  $n$-hadron inclusive cross-section. 

\begin{figure}
\hspace{1cm}
\resizebox{1.75in}{1.25in}{\includegraphics[0.5in,0.5in][5in,3.5in]{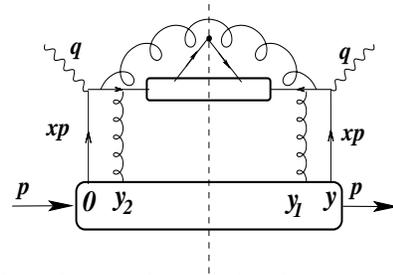}}
\caption{A typical higher twist contribution used to compute the modification of the fragmentation 
function in medium.}
\label{fig2}
\end{figure}

One assumes the 
hierarchy of scales $E >> Q >> \mu$ and applies this to the computation of multiple Feynman 
diagrams such as the one in Fig.~\ref{fig2}, this diagram represents 
the process of a hard virtual quark produced in a hard collision, which radiates 
a gluon and then scatters off a soft medium gluon with transverse momentum 
$q_\perp \sim \mu$ prior to exiting the medium and fragmenting into hadrons. 
At a given order, there exist various other contributions which involve scattering of 
the initial quark off the soft gluon field prior to radiation as well as scattering of the 
radiated gluon itself. All such contributions are combined coherently to calculate the 
modification to the fragmentation function directly.  

The hierarchy of scales allows 
one to use the collinear approximation to factorize the fragmentation function and its 
modification from the hard scattering cross section. 
Thus, even though such a modified 
fragmentation function is derived in DIS, it may be generalized to the kinematics of a 
heavy-ion collision. 
Diagrams where the outgoing parton scatters off the medium gluons, 
such as those in Fig.~\ref{fig2},
produce a medium dependent additive contribution to the vacuum fragmentation function, 
which may be expressed as, 

\bea
\mbox{}\!\!\!\!\!\!\!\!\!\D D_i(z,\mu_f^2) &=& \int_{0}^{\mu_f^2} \frac{dk_{\perp}^2}{k_{\perp}^2} 
\frac{\A_s}{2\pi} \left[ \int_{z_h}^1 \frac{dx}{x} 
\sum_{j=q,g} \right.  \\ 
\ata \left. \left\{ \D P_{i \ra j} (x,x_B,x_L,k_\perp^2) 
D_j^{h} \left(\frac{z_h}{x}, \mu_f^2 \right) 
\right\} \right]. \nn
\label{med_mod}
\eea
\nt
In the above equation,  $\D P_{i\ra j}$ 
represents the medium modified splitting function of parton $i$ into $j$ 
where a momentum fraction $x$ is left in parton $j$. 
The argument $x_L = k_\perp^2/(2P^-p^+ x(1-x))$ is a momentum fraction defined such that $x_L P^-$ is 
the formation time of the radiated parton \footnote{Throughout the HT portion of this work, four-vectors  
will often be referred to using the light cone convention where $x^\pm = (x^0 \pm x^3)/\sqrt{2}$. For the 
higher-twist scheme, often, $x^+ = ( x^0 + x^3 )/2$ and $x^- = x^0 - x^3$.}, where 
the radiated gluon or quark carries away a transverse momentum $k_\perp$, $P^-$ is 
the incoming momentum of a nucleon in the nucleus and $p$ is the momentum of the 
virtual photon. 
The scale $\mu_f$ refers to the hard scale of the process.
The medium modified splitting functions may be expressed as a product 
of the vacuum splitting function $P_{i \ra j}$ and a medium dependent factor, 

\bea
\D \hat{P}_{i\ra j} &=& P_{i \ra j}(x)  \frac{C_A 2\pi \A_s   T^A_{qg} (x_B,x_L)}{(k_\perp^2 + 
\lc q_\perp^2 \rc)  N_c f_q^A(x_B)} .  \label{mod_split}
\eea
\nt
Where, $C_A,N_c$ represent the adjoint Casimir and the number of colours. 
The mean 
transverse momentum of the soft gluons is represented by the factor $\lc q_\perp^2 \rc$.  
The term $T^A_{qg}$ represents the quark gluon correlation in the
nuclear medium, and depends on the four point correlator,  
\bea 
\mbox{}\!\!\!\!\!\!\!\!\!\!\!\!\!\!\!\mbox{} & & 
\lc  P | \psibar (0)  \g^- F_\sg^- ( y_2 ) {F^{-}}^\sg (y_1 ) \psi (y) | P \rc  \nn \\
\mbox{}\!\!\!\!\!\!\!\!\!\!\!\!\!\mbox{}  &\sim&  
C \lc p_1| \psibar (0)  \g^- \psi (y) | p_1 \rc 
\lc p_2 |  F_\sg^-\!( y_2 ) {F^{-}}^\sg \!(y_1 )| p_2 \rc. \label{FF}
\eea
\nt
Where, $F_\sg^- ( y_2 )$ and $ {F^{-}}^\sg (y_1 )$ represent gluon field 
operators at the locations $y_1,y_2$ and $\psi(y)$ represents the quark field operator. 
The above correlation function cannot be calculated from first principles without making 
assumptions regarding the structure of the medium. 
The only assumption 
made is that that the colour correlation length is small.
As a result, one may factorize 
the four point function into two separate structure functions, one for the original parton produced in the 
hard scattering [this is a quark in Eq.~\eqref{FF}] and one for the soft gluon off which 
the parton scatters in the final state. 

While in media with short distance color correlation 
lengths such as the atomic nucleus or a QGP with a large Debye mass, this factorization 
may be generally thought to be true, it may fail at very large jet energies where saturation effects become important. 
It should also be pointed out that the factorization assumption above falls in the same class as the 
assumption of independent scattering centers as assumed in the ASW or GLV scheme. 
In the application of this formalism to RHIC data we have assumed that the jet energies 
are not high enough for the onset of saturation effects. Another scenario where the above 
factorization may not hold is if there were long distance color correlations in the QGP, which 
have been assumed to be absent. If such long distance correlations were present then one would 
have to resort to the definition of more general multi-particle operators [such as the first line in Eq.~\eqref{FF}]
and parametrize these in comparison with experimental data.

The entire phenomenology of the medium is incorporated as a model for the expectation of the 
second set of operators in Eq.~\eqref{FF}. 
This may be characterized in terms of the well known medium transport coefficient 
$\hat{q}(\zeta)$, at location $\zeta$, where, 
\bea
\hat{q}(\zeta) &=& \frac{4 \pi^2 \A_s C_R}{N_c^2 - 1} 
\int \frac{d \xi^+}{2 \pi}  \frac{ d^2 \xi_\perp d^2k_\perp}{(2\pi)^2}  \label{qhat} \\
\ata \exp \left[ i \frac{q_\perp^2}{2 p^+}  \xi^+ - i \vp_\perp \x \vec{\xi}_\perp  \right] \nn \\
\ata  \lc F^{-,}_\sg (\zeta + \xi^+/2 , \vec{\xi}_\perp/2) 
F^{\sg -} (\zeta - \xi^+/2 , -\vec{\xi}_\perp/2 ) \rc.  \nn
\eea
The Casimir $C_R$ depends on the representation of the probe. 
The transport coefficient is normalized by fitting to one data point and a model such as a Woods-Saxon 
distribution for cold matter or 3-D hydrodynamical evolution for hot nuclear matter 
is invoked for its variation with space-time location. The expectation $\lc \,\, \rc$ 
is meant to be taken in the medium under consideration. Any space time dependence 
is essentially included in the implied expectation. 

Closer inspection of Eq.~\eqref{qhat} reveals that it is a function of the jet energy $p^+$. 
Note that $p^+ $ is not integrated out. The actual dependence on $p^+$ depends on the medium 
in question. In the case of confined nuclear media, or a quark gluon plasma, the dependence is 
logarithmic. 
There is also a logarithmic dependence on the virtuality 
of the jet which sets in due to radiative corrections to the definition in Eq.~\eqref{qhat}.
Also, as demonstrated in Ref.~\cite{Majumder:2006wi}, $\hat{q}$ may even possess a tensorial 
structure if the medium is not isotropic.
In the calculations of the current manuscript, both the dependence on the energy and 
virtuality of the jet will be ignored. The medium will be assumed to be isotropic. The 
values of $\hat{q}$ quoted should thus be 
considered as approximations to the full functional form.

Unlike the remaining formalisms, 
the HT approach is set up to directly calculate the medium modified fragmentation function 
and as a result the final distribution of hadrons. This modification to the distribution includes both contributions 
coming solely from the medium and those which involve interference between medium induced and 
vacuum radiation.  
The determined constant, $\hat{q}$, may be used to calculate
the average energy loss encountered by a jet. Other advantages of this approach 
include a functional difference between the quark and gluon energy loss kernels, \tie, 
the difference between the modification as encountered by a quark jet and a gluon jet is not 
merely assumed to be a ratio of Casimirs, but depends strongly on the different splitting and 
fragmentation functions.  This formalism, offers by far the most 
straightforward generalization to multi-particle correlations~\cite{Majumder:2004wh} and 
their modification in the medium.

A disadvantage of this approach at the current state of approximation 
(similar to the GLV and the ASW but different from the AMY approach) 
is the negligence of the quark structure function in the medium: as a result, collisions with the medium may not change the
flavor of the jet parton, however this may continue to occur through the splitting kernels. Yet another disadvantage 
is the restriction to single scattering followed by single radiation in the medium, which makes this formalism more 
appropriate to thin media. This is partially improved by converting Eq.~\eqref{med_mod} 
to an evolution equation as in \cite{Guo:2000nz} which describes the virtuality evolution of the probe in the medium.

As this formalism is originally cast in cold nuclear matter, the applicability of the formalism only 
depends on there being a short distance color correlation length in the medium. As a result, it may 
be used to describe both confined and deconfined matter with the inclusion of an Ansatz for the 
variation of $\hat{q}$ with an intensive property of the medium such as energy density $\ep$, 
entropy density $s$ or the temperature $T$ and baryon chemical potential $\mu_B$.

\subsection{Opacity expansion: Quenching weights formalism}


The path integral approach for the energy loss of  a hard jet propagating 
in a colored medium was first introduced in Ref.~\cite{Zakharov:1996fv}.
It was later demonstrated to be equivalent to the well known BDMPS 
approach~\cite{Baier:1996kr,Baier:1996sk} in the limit of multiple scatterings~\cite{Zakharov:2004vm}. 
The current, most widespread, variant of this 
approach developed by numerous authors~\cite{Salgado:2003gb,Armesto:2003jh} is often referred to as the 
Armesto-Salgado-Wiedemann (ASW) approach. In this scheme, one 
incorporates the effect of multiple scattering of the incoming  and 
outgoing partons in terms of a path integral over a path ordered 
Wilson line~\cite{Wiedemann:2000ez}.   

This formalism assumes a model for 
the medium as an assembly of Debye screened heavy scattering centers which are well separated 
in the sense that the mean free path of a jet 
$\lambda \gg 1/\mu$ the colour screening length of the medium~\cite{Gyulassy:1993hr} . 
The opacity of the medium $\bar{n}$ quantifies the number of scattering centers seen  by a 
jet as it passes through the medium, \tie, $\bar{n} = L/\lambda$, where $L$ is the 
thickness of the medium.
A hard, almost on shell,  parton traversing such a medium will 
engender multiple transverse scatterings of  order $\mu \ll p^+$. It will, in the 
process, split into an outgoing parton and a radiated gluon which will also scatter 
multiply in the medium. 
The propagation of the incoming (outgoing) partons as well as that of the radiated gluon 
in this background colour field may be expressed in terms of  effective Green's functions 
[$G(\vecr_\perp,z ; \vecr_{\perp}\,\p, z\p)$ (for quark or gluon)] which obey the obvious 
Dyson-Schwinger 
equation, 

\bea
\mbox{}\!\!\!\!\!\!\!\!\!\!\!\!\!&& G(\vecr_\perp,z ; \vec{r\p}_\perp, z\p) =G_0(\vecr_\perp,z ; \vec{r\p}_{\perp}, z\p) \nn \\ 
\mbox{}\!\!\!\!\!\!\!\!\!\!\!\!\!&-&i \int_z^{z\p}\!\!\!\!\!\! d \zeta \!\!\int \!\!\! d^2\vx 
G_0(\vecr_\perp, z; \vx, \zeta) A_0(\vx,\zeta) G (\vx, \zeta; \vec{r\p}_{\perp}, z\p)  \label{ASW1},
\eea
\nt
where, $G_0$ is the free Green's function and $A_0$ represents the color potential of  the medium. 
The solution for the above interacting Green's function involves a path ordered Wilson line  which 
follows the potential from the location $[\vecr_{\perp}(z\p), z\p]$ to $[\vecr_{\perp}(z), z]$. Expanding 
the expression for the radiation cross section to order $A_0^{2n}$ corresponds to an expansion 
up to $n^{th}$ order in opacity.

Taking the high energy limit and the soft radiation approximation ($x<<1$), one focuses on isolating 
the leading behavior in $x$ that arises from the large number of  interference diagrams at a given 
order of opacity. As a result of the approximations made, one recovers the BDMPS condition that 
the leading behavior in $x$ is contained solely in gluon re-scattering diagrams.  This results in the 
expression for the inclusive energy distribution for gluon radiation off an in-medium produced parton as~\cite{Wiedemann:2000tf},

\bea
\w \frac{dI}{d\w} &=& \frac{\A_s C_R}{(2\pi)^2 \w^2} 2 {\rm Re}\!\!\! \int\limits_{\zeta_0}^\infty \!\!\!d y_l  
\!\!\int\limits_{y_l}^{\infty} \!\!\! d \bar{y}_l 
\!\!\int \!\!\! d\vec{u}\!\!\!\! \int\limits_0^{\chi x p^+} \!\!\! d \vec{k} e^{- i \vk \x \vec{u} - \hf \int d\zeta n(\zeta) \sg(\vec{u}) } \nn \\
&\times& \!\!\!\!\frac{\prt^2}{\prt y \prt u}\!\!\!\!\!\!\!\!\int\limits_{\vec{y}=0=\vecr(y_l)}^{\vec{u}=\vecr(\bar{y})}\!\!\!\!\! \mathcal{D}r 
e^{i\int d\zeta \frac{\w}{2} \left( |\dot{\vecr}|^2  - \frac{n(\zeta) \sg(\vecr) }{i \w}\right)}, \label{ASW2}
\eea 
\nt
where, as always, $k_\perp$ is the transverse momentum of 
the radiated gluon with energy $\w$ and $\chi$ is a factor that introduces the kinematic bound. 
The vectors $\vec{y}$ and $\vec{u}$ represent
the transverse locations of the emission of the gluon in 
the amplitude and the complex conjugate whereas $y_l$ and $\bar{y}_l$
represent the longitudinal positions.  The density of 
scatterers in the medium at location $\zeta$ is $n(\zeta)$ and
the scattering cross section is $\sg(r)$. In this form, the 
opacity is obtained as $\int n(\zeta) d \zeta$ over the extent of the medium.
The Casimir $C_R$ depends on the representation of the 
jet parton.

Exact analytical expressions for the gluon radiation 
intensity distribution are rather involved and only yield simple expressions in 
certain special circumstances.
Numerical implementations of this scheme have focused on
two separate regimes. In one case, $\sg(r)$ is replaced with a
dipole form $\hat{q}r^2/n(\zeta)$ and one solves the harmonic oscillator like
path integral. This corresponds to the case of multiple 
soft scatterings of the hard probe. 
In the limit of a static medium with a very large length, 
one obtains the simple form for the radiation distribution, 

\bea
\w \frac{dI}{d\w} \simeq \frac{2 \A_s C_R}{\pi}  \left\{ \begin{array}{lcr} 
\sqrt{\frac{\w_c}{2\w}} \hspace{0.5cm}  &  \mbox{for} & \w < \w_c,  \\
\frac{1}{12} \left( \frac{\w}{\w_c} \right)^2 & \mbox{for} &  \w > \w_c. 
\end{array} \right. \label{omega_c}
\eea
Where $\w_c = \int d \zeta \zeta \hat{q} (\zeta)$ is called the characteristic frequency of the radiation. 
Up to constant factors, this is equal to 
mean energy lost in the medium ($\lc E \rc$) \tie, $\w_c \simeq  2 \lc E \rc / (\A_s C_R) $. 
In the other extreme, one expands 
the exponent as a series in $n\sigma$; keeping 
only the leading order term corresponds to the picture of gluon 
radiation associated with a single scattering. 
In this case, the gluon emission intensity distribution has been found to 
be rather similar, once scaled with the characteristic 
frequency appropriate for this situation. For dynamical medium of finite extent, the characteristic 
frequency and the overall mean transverse momentum gained by the jet $\lc \hat{q} L \rc $ will have to be 
estimated based on an Ansatz for the space time distribution of the transport parameter $\hat{q}$.  

Due to the soft limit \tie, $\w \ra 0$ used, multiple gluon emissions are required 
for a substantial amount of 
energy loss. Each such emission at a given opacity is assumed independent and a 
probabilistic scheme is set up, wherein, the jet loses an energy fraction $\D E$ in 
$n$ tries with a Poisson distribution~\cite{Salgado:2003gb}, 

\bea
P_n(\D E) = \frac{e^{-\lc I \rc} }{n!} \Pi_{i=1}^n \llb \int d \w_i \frac{dI}{d\w_i}  \lrb 
\kd(\D E - \sum_{i=1}^{n} \w_i  )  . \label{ASW3}
\eea

\nt
where, $\lc I \rc$ is the mean number of gluons radiated per coherent interaction set.
Summing over $n$ gives the probability  $P(\D E)$ for  an incident jet to lose 
a momentum fraction $\D E$ due to its passage through the medium. 
This is then used to model a medium 
modified fragmentation function, by shifting the energy fraction available to 
produce a hadron (as well as accounting for the phase space available after energy loss), 

\bea
\tilde{D}(z,Q^2) = \int_0^{1} d \D E  P(\D E) \frac{D\left( \frac{z}{1-\D E},Q^2\right)}{1-\D E}. \label{ASW4}
\eea
\nt
The above, modified fragmentation function is then used in a factorized formalism as in Eq.~\eqref{basic_cross} 
to calculate the final hadronic spectrum. 

In marked contrast to other approaches, this scheme presents the advantage of easy interpolation between the 
cases of few hard scatterings and multiple soft scatterings and is thus applicable to both thin and thick media. 
The inclusion of the zero opacity term makes this the 
only formalism, to date, which includes interference between vacuum radiation and radiation induced by 
multiple soft scattering in the medium.  
It suffers from the disadvantage of having approximated 
the medium in terms of heavy static scattering centers. As a result, elastic energy loss is vanishing in this scheme. 
As the formalism is setup to calculate the energy loss probability of the leading hard parton, estimation of the change in the 
distribution of final associated (sub-leading) hadrons or partons is not straightforward. 

Along with the HT formalism, this approach 
also neglects any flavor changing scatterings in the medium. Also similar with HT is the treatment of both confined and 
deconfined matter on the same footing: one essentially makes an Ansatz for the variation of $\hat{q}$ with an intensive 
variable of the medium \eg, $\ep, s, T, \mu_B$.


\subsection{Finite temperature field theory formalism}


In this scheme, often referred to as the Arnold-Moore-Yaffe (AMY) approach, the energy loss
of hard jets is considered in an extended medium in equilibrium at asymptotically high temperature $T \ra \infty$. Owing
to asymptotic freedom  the coupling constant $g \ra 0$ at such high temperatures, and a power counting scheme emerges
from the ability to identify  a hierarchy of parametrically separated scales $T >> gT >> g^2 T$ \etc In this limit, it
then becomes possible to construct an effective field theory of soft modes, \tie, $p \sim gT$ by summing contributions from
hard loops with $p \sim T$, into effective propagators and vertices~\cite{Braaten:1989kk}.

\begin{figure}
\hspace{0cm} \resizebox{1.5in}{1.25in}{\includegraphics[2.5in,0.0in][7in,3in]{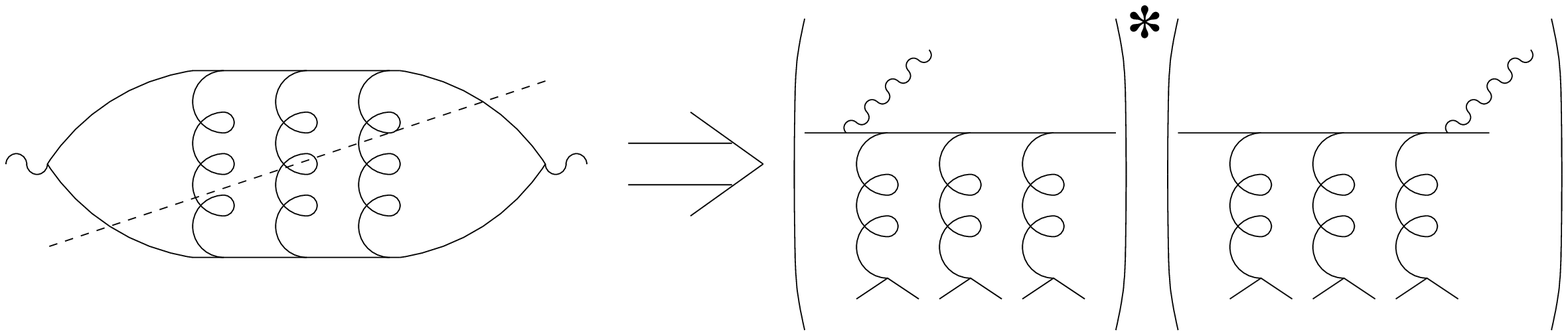}}
\caption{A ladder diagram evaluated in the AMY formalism} \label{fig3}
\end{figure}

One assumes a hard on-shell parton, with energy several times that of the temperature, traversing such a medium,
undergoing soft scatterings with momentum transfers $\sim gT$ off other hard partons in the medium. Such soft
scatterings induce collinear radiation from the parton, with a transverse momentum of the order of $g T$. The
formation time for such collinear radiation $\sim 1/(g^2T) $ is of the same order of magnitude as the mean free time
between soft scatterings~\cite{Arnold:2001ba}. As a result, multiple scatterings of the incoming (outgoing) parton and
the radiated gluon need to be considered to get the leading order gluon radiation rate. One essentially calculates the
imaginary parts of infinite order ladder diagrams such as those shown in Fig.~\ref{fig3}; this is done by means of
integral equations~\cite{Arnold:2002ja}.

The imaginary parts of such ladder diagrams yield  the $1\ra2$ decay rates of a hard parton into a radiated gluon and
another parton. These decay rates are then used to evolve hard quark and gluon distributions from the initial hard
collisions, when they are formed, to the time when they exit the medium, by means of a set of coupled Fokker-Planck
like equations for quarks, anti-quarks and gluons~\cite{Qin:2007zz, Jeon:2003gi, Turbide:2005fk}, which may be written
schematically as,
\begin{eqnarray}
\label{FP-eq} \frac{dP_j(p,t)}{dt} &=& \sum_{ab} \int dk \left[P_a(p+k,t) \frac{d\Gamma^{a}_{jb}(p+k,p,t)}{dk dt}
\right.\nonumber\\ && \left. - P_j(p,t)\frac{d\Gamma^{j}_{ab}(p,k,t)}{dk dt}\right].
\end{eqnarray}
In the above equation, $j = q, \bar{q}, g$, and we sum over all relevant partonic processes for each evolution
equation. In contrast to all other schemes, this approach also includes the absorption of thermal gluons as well as
quark anti-quark pair annihilation and creation.

The initial jet distributions are taken from a factorized hard scattering cross section as in Eq.~\eqref{basic_cross}.
In the limit of single scattering, these rates may be taken directly from the corresponding Gunion-Bertsch cross
sections~\cite{Gunion:1981qs} for an on-shell parton to radiate a gluon on soft scattering with another in-medium
parton.

The final hadron spectrum at high $p_T$ is obtained by the fragmentation of jets in the vacuum after their passing
through the medium. In this approach, one calculates the medium modified fragmentation function by convoluting the
vacuum fragmentation functions with the hard parton distributions, at exit, to produce the final hadronic
spectrum~\cite{Turbide:2005fk},
\begin{eqnarray}
\label{AMY2_FF} \tilde{D}^h_{j}(z,\vec{r}_\bot, \phi) \!&=&\!\! \sum_{j'} \!\int\! dp_{j'} \frac{z'}{z} D^h_{j'}(z')
P(p_{j'}|p_j,\vec{r}_\bot, \phi). \ \ \ \ \ \
\end{eqnarray}
where the sum over $j'$ is the sum over all parton species. The two momentum fractions are $z=p_h/p_j$ and
$z'=p_h/p_{j'}$, where $p_j$ and $p_{j'}$ are the momenta of the hard partons immediately after the hard scattering
and prior to exit from the medium, and $p_h$ is the final hadron momentum. $P(p_{j'}|p_j,\vec{r}_\bot, \phi)$
represents the solution to Eq.~\eqref{FP-eq}, which is the probability of obtaining a given parton $j'$ with momentum
$p_{j'}$ when the initial condition is a parton $j$ with momentum $p_j$. The above integral depends implicitly on the
path taken by the parton and the medium profile along that path, which in turn depends on the location of the origin
$\vec{r}_{\bot}$ of the jet, its propagation angle $\phi$ with respect to the reaction plane. Therefore, one must
convolve the above expression over all transverse positions $\vec{r}_{\bot}$ and directions $\phi$.

The use of an effective theory for the description of the medium and the propagation of the jet, makes this approach
considerably more systematic than the two previous approaches: both the properties of the jet and the medium are
described using the same hierarchy of scales. It remains the only approach to date which naturally includes partonic
feedback from the medium, \tie, processes where a thermal quark or gluon may be absorbed by the hard jet
\footnote{While an attempt to include such effects in the higher twist formalism have been made in
Ref.~\cite{Wang:2001cs}, these remain as phenomenological extentions and have not been included in this manuscript.}.
In contrast to ASW and HT, this approach also includes flavor changing interactions in the medium. Elastic energy loss
may also be incorporated within the same basic formalism~\cite{Qin:2007rn}. Note that AMY assumes a thermalized
partonic medium and neglects the quenching of jets in the confined sector. In addition, interference between medium
and vacuum radiations is not yet considered.

The use of HTL effective theory to describe both the jet propagation in
the medium and the properties of the medium
itself does suffer from one caveat: this scheme approximates the bulk
structure of the medium as a weakly
coupled plasma of quarks and gluons. The perturbative estimates of the
energy density ($\epsilon$) differs from the $\epsilon(T)$ obtained
from lattice calculations (at $3 T_c \geq T \gtrsim T_c$). The
$\eta/s$ required to
reproduce the observed magnitude of elliptic flow in viscous fluid
dynamical simulations is at least a factor of
two lower than perturbative results. The application of such a scheme to
the modification of hard
jets involves an aspect of phenomenology where the coupling constant is
used as a fit parameter.


\subsection{Geometry and Discussion of the different Schemes}


As mentioned previously, all parameters of our hydrodynamic evolution \cite{Nonaka:2006yn}
have been fixed by a fit to the soft sector (elliptic flow, pseudo-rapidity
distributions and low-$p_T$ single particle spectra), therefore
providing us with a fully determined 
medium evolution for the hard probes to propagate through.
The hydrodynamic calculation provides a time-evolution of the temperature, energy-density, flow velocity and QGP to hadron gas
fraction within all hydrodynamic cells composing the medium through which the hard probes
evolve. The incorporation of this information within the different jet energy-loss schemes is 
described in the following subsections.

The  mean  impact parameters for the different evolution sets have been set to $b=$2.4, 4.5, 6.3, 7.5 fm, 
corresponding
to 0-5, 10-15, 15-20, 20-30 \% centrality, respectively.
These values were estimated via the number of 
nucleon-nucleon binary collisions and the number of participant nucleons 
in Ref.~\cite{Adler:2003cb}. In this work, the focus will lie on the two extreme centrality 
bins in the list above: the 0-5\% bin and the 20-30\% bin. All of the RFD calculations utilized here have 
an initial thermalization time of $\tau_0 = 0.6$fm/c. Any values of parameters, such as $\hat{q}$, which are dependent on the 
bulk properties of the medium will be quoted at this time.

All three energy loss schemes are sensitive to certain bulk properties of the 
evolving matter: while in the case of the AMY formalism this is decidedly the temperature, the 
relation between $\hat{q}$ and the intensive variables of the medium in the HT and ASW formalisms is unspecified. 
Traditionally, the $\hat{q}$ in the ASW scheme has been related to the  the energy density 
$\epsilon$ via  $\ep^{3/4}$
while the 
$\hat{q}$  in the HT scheme has been scaled either with the temperature $T$ via $T^3$ or
the entropy density $s$ of the local medium. 
In the analysis presented in this paper we maintain this methodology, however, some surprising results of 
scaling the ASW \qhat with $T^3$ and the HT \qhat with $\ep^{3/4}$ will also be presented.

\begin{figure}[tbh]
\includegraphics[width=\columnwidth]{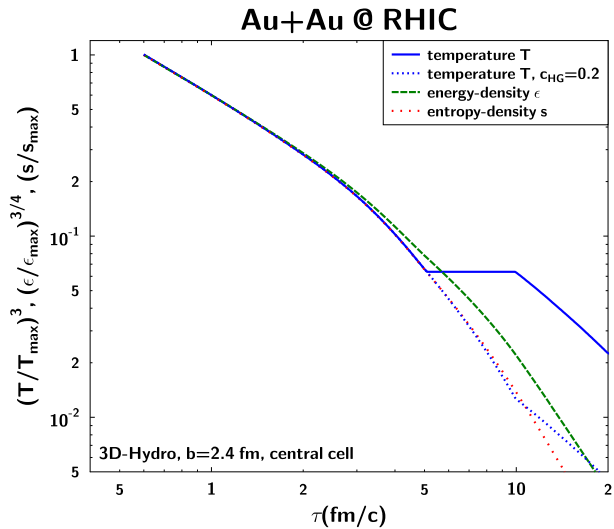}
\caption{(Color online) Time evolution of temperature $T^3$, energy-density $\epsilon^{3/4}$  and
entropy density $s$ as a function of time $\tau$ in the central cell of the hydrodynamic
evolution for Au+Au collisions at RHIC. All curves are normalized to their maximum values
at $\tau=0.6$~fm/c.}
\label{tevol}
\end{figure}

The scaling of $\hat{q}$ with $\ep^{3/4}$,
$T^3$ or $s$ will by construction yield identical results for a QGP with an ideal gas
equation of state: $\epsilon = 3 p$. However,
for a more realistic non-ideal equation of state as used in our hydrodynamic calculation,
the value of $\hat{q}$ will be affected by the choice of scaling variable, in particular
if energy-loss persists to temperatures below $T_C$. Figure~\ref{tevol}, investigates
the deviations from the ideal gas scaling by plotting the normalized time evolution
of temperature $T^3$, energy-density $\epsilon^{3/4}$  and
entropy density $s$. As can be seen after 2 fm/c the curves start to deviate from the
ideal gas power law behavior and start to show differences for times later than $\tau=3$~fm/c.
The first order phase-transition contained in our equation of state results in a striking
difference between the temperature and the energy- or entropy-density scaling at the critical
temperature and below. We note that the {\em proper} scaling law for $\hat{q}$ is {\it a priori} 
not known, even though we see no reason why it should not be calculable in QCD.

\subsubsection{Higher Twist}


In the preceding section, the medium modification of the final fragmentation function 
in the Higher Twist formalism was shown to be dependent on the transport coefficient 
\qhat [see Eq. \eqref{qhat}]. In the evolving system formed in the collision of two nuclei, this transport coefficient has 
both a space and time dependence [i.e., \qhat$(x,y,z,\tau)$]. Phenomenologically, this dependence is taken to scale with 
some intensive variable of the medium, in this case, the dimensionally equivalent quantities of 
$T^3$ or the entropy density $s$, i.e.,
\bea
\hat{q}(x,y,z,\tau) &=& \hat{q}_{0}  
\frac{  \gamma_\perp (x,y,z,\tau) T^3(x,y,z,\tau) }{T_0^3} \label{qhat_pres} \\
\mbx \times && \!\!\!\!\!\!\!\!\!\!\!\!\left[ R(x,y,z,\tau) + c_{HG} \left\{  1 - R(x,y,z,\tau) \right\}  \right] , \nn
\eea
\nt
where, $T(x,y,z,\tau)$, $\gamma_\perp (x,y,z,\tau)$ and $R(x,y,z,\tau)$ represent the temperature, 
flow transverse to the jet and the volume fraction in the plasma phase at the space-time 
point $x,y,z,\tau$. It is this information that 
is extracted from the RFD simulation. Though the RFD simulations start at $\tau= 0.6$ fm/c, the 
values of $T$ and $\gamma_\perp $ at $\tau=0.6$ fm/c are extrapolated as constants to $\tau =0$ which 
represents the time of the initial hard scatterings (the effect 
of different extrapolation schemes involving linearly rising or dropping values of \qhat as $\tau \ra 0$ 
has been found to be rather small and will not be studied in this effort).
The factors $\hat{q}_0,T_0$ represent the maximum 
$\hat{q}$ and temperature achieved in the simulation; in this particular version of RFD, $T_0 = 0.405$ GeV and 
$\hat{q}_0$ is a fit parameter adjusted to fit one data point of the $R_{AA}$, at one centrality. 

The factor $c_{HG}$ may be interpreted in two ways. In essence, it accounts for the fact that the quenching in 
the hadronic phase may not be as effective as that in the partonic phase at the same temperature. Since the 
entropy density in a given phase is proportional to $T^3$ with the constant of proportionality demonstrating 
a weak dependence on temperature, $c_{HG}$ may be tuned to convert the scaling of \qhat with $T^3$ 
into a scaling with $s$. This is approximately achieved with a $c_{HG} \sim 0.2$,
as can be seen in Fig.~\ref{tevol}. Unless specified 
otherwise, this is the value used for $c_{HG}$ in 
all the plots in this paper. Thus $c_{HG}$ is not a fit parameter and is not tuned to fit any experimental 
data point. It has only three possible choices of $c_{HG} =0$ which corresponds to no quenching in the hadronic
phase,  $c_{HG} =1$ which corresponds to exact scaling of \qhat with $T^3$ and $c_{HG} = 0.2$ which 
corresponds to approximate scaling of \qhat with $s$.

Given a choice of $c_{HG}$ and the overall fit parameter $\hat{q}_0$, we use Eq.~\eqref{med_mod} to 
calculate a medium modified fragmentation function; then Eq.~\eqref{basic_cross} is used to compute the
total cross section and the nuclear modification factor $R_{AA}$. 
The overall fit parameter $\hat{q}_0$ is tuned to fit one experimental data point, 
at one centrality and $p_T$. For the current effort, the fit parameter is set by requiring that the 
$R_{AA}$ at $p_T=10$ GeV in the most central event ($0$-$5\%$ centrality) is $0.2$. 
With the value of $\hat{q}_0$  and $c_{HG}$ fixed, the variation of $R_{AA}$ as a function of 
$p_T$ (integrated over or with respect to the angle with the reaction plane) and centrality of the collision are predictions.


\subsubsection{ASW}


As in the previous case, we have to formulate the energy loss problem for a dynamical medium in which the transport coefficient $\hat{q}$ acquires a space and time dependence. As done in previous calculations within the ASW formalism, we use a scaling with the local energy density $\epsilon^{3/4}$ along the path $\xi = (x(\tau), y(\tau), z(\tau), \tau)$ of a parton as

\begin{equation}
\label{E-qhat}
\hat{q}(\xi) = K \cdot 2 \cdot \epsilon^{3/4}(\xi).
\end{equation}

This scaling of $\hat{q}$ is assumed to be valid in both the partonic and the hadronic phase. The precise form of the path $\xi$ is determined once the hard initial vertex $(x_0,y_0)$ in the transverse plane, the outgoing parton rapidity $\eta$ and the angle of the parton with the reaction plane $\phi$ is specified. The parameter $K$ in Eq.~(\ref{E-qhat}) is regarded as a parameter to account for the uncertainty in the selection of $\alpha_s$ and possible non-perturbative effects increasing the quenching power of the medium (see discussion in \cite{Renk:2006pk}).

Given this spacetime dependence of the transport coefficient along a parton trajectory, the energy loss probability distribution can be computed from the two line integrals

\begin{equation}
\label{E-omega}
\omega_c({\bf r_0}, \phi) = \int_0^\infty \negthickspace d \xi \xi \hat{q}(\xi) \quad  \text{and} \quad \langle\hat{q}L\rangle ({\bf r_0}, \phi) = \int_0^\infty \negthickspace d \xi \hat{q}(\xi).
\end{equation}

Here, $\omega_c$ is the characteristic gluon frequency, setting the scale of the energy loss probability distribution (see expression~\ref{omega_c}) and $\langle \hat{q} L\rangle$ is a measure of the path-length, weighted by the local quenching power.
Analogously to the overall fit parameter $\hat{q}_0$ in the HT case, the parameter $K$ is fit one data point of the $R_{AA}$, at one centrality. 

For times prior to $\tau = 0.6$ fm/c, i.e. the starting point of the RFD simulation, we neglect any medium effects, i.e. assume $\hat{q} = 0$. Note that for a purely radiative energy loss model where the average energy loss grows quadratically with pathlength in a constant medium the effect of initial time dynamics is systematically suppressed and no strong dependence of the energy loss on variations of the initial time is observed.

Using a dynamical scaling law \cite{Salgado:2002cd}, $\omega_c$ and $\langle \hat{q} \rangle$ can then be mapped onto a static equivalent scenario. Using the relation $R=2\omega_c^2/\langle\hat{q}L\rangle$ as an input, we then determine $P(\Delta E)$ using the numerical results from \cite{Salgado:2003gb} and compute the medium-modified fragmentation function from Eq.~(\ref{ASW4}). The resulting expression (valid for a single path) must then be averaged over the whole geometry with a weight corresponding to the probability of finding an initial hard vertex at $x_0,y_0$.


\subsubsection{AMY}


The strength of the transition rate in pQCD is controlled by the strong coupling constant $\alpha_s
(T)$, temperature $T$ and the flow parameter $\vec{\beta}$ (the velocity of thermal medium) relative to the jet's
path. The value for the coupling constant used (along with the assumption of a thermalized partonic medium) may be
related to the transport coefficient $\hat{q}$ as derived from computations in HT or ASW by either a direct
computation of the operator product in Eq.~\eqref{qhat}, or a computation of the mean transverse momentum squared per
unit length as gained by a jet which propagates through the medium without radiation.

In a 3D expanding medium, the transition rate is first evaluated in the local frame of the thermal medium, then
boosted into the laboratory frame,
\begin{eqnarray}
\left.\frac{d\Gamma(p,k,t)}{dk dt}\right|_{\rm lab} &=& (1-\vec{v}_j\cdot\vec{\beta})
\left.\frac{d\Gamma(p_0,k_0,t_0)}{dk_0dt_0}\right|_{\rm local}, \ \ \ \ \ \
\end{eqnarray}
where $k_0=k(1-\vec{v}_j\cdot\vec{\beta})/\sqrt{1-\beta^2}$ and $t_0=t\sqrt{1-\beta^2}$ are momentum and the proper
time in the local frame. As jets propagate in the medium, the temperature and the flow parameters depend on the time
and the positions of jets, and the 3D hydrodynamical calculation \cite{Nonaka:2006yn} is utilized to determine the
temperature and flow profiles. The energy loss mechanism is applied at time $\tau_0=0.6$~fm/c, when the medium reaches
thermal equilibrium, and switched off when the medium reaches the hadronic phase.


\section{Application to RHIC Data} \label{results}


In the preceding sections, a description of the theoretical setup underlying each of the three schemes as 
well as the phenomenological connection between them and the RFD simulations was 
expounded upon. In this section, we present the results of our numerical calculations. The 
primary quantity of interest will be the nuclear modification factor ($R_{AA}$) defined as 

\begin{eqnarray}
R_{AA} &=& \frac{ \frac{d \sigma^{AA}(b_{min}, b_{max})}{dy d^2 p_T} }
{\int_{b_{min}}^{b_{max}} d^2 b T_{AA} (b) \frac{d \sigma^{pp}(p_T,y)}{dy d^2 p_T}  }, \label{raa} \\
&\simeq& \frac{  \frac{ d \sigma^{AA} (\lc b \rc)}{d^2b dy d^2 p_T}}{T_{AA} (\lc b \rc) \frac{d \sigma^{pp}(p_T,y)}{dy d^2 p_T}}, \nn 
\end{eqnarray} 
where, $T_{AA}$ represents the nuclear overlap function which is proportional to the number of binary collisions at the 
mean impact parameter $\lc b \rc$. The mean impact parameter for a given range of centrality is 
essentially set by the RFD simulation used to calculate the soft observables. The $R_{AA}$ is calculated both integrated 
as defined above or as function of the angle with respect to the reaction plane. 

The range of $p_T$ of the detected hadron is set high enough for the applicability of pQCD. 
In this paper, the lower bound is 
set at $p_T = 6$ GeV. This choice is essentially dictated by the regime where recombination~\cite{Fries:2003vb} begins to 
contribute to the yield. The upper limit is set at $p_T = 20$ GeV which represents the highest $p_T$ for which experimental 
data exist. The focus in this paper will essentially be on two different centralities of $\lc b\rc =2.4$ fm which represents 
the rather central collisions with a centrality in the range from $0-6\%$ and a somewhat more peripheral event with a 
$\lc b\rc =7.5$ fm which corresponds to the $20-30\%$ range of centrality.

\begin{figure}[tbhp]
\includegraphics[width=\columnwidth]{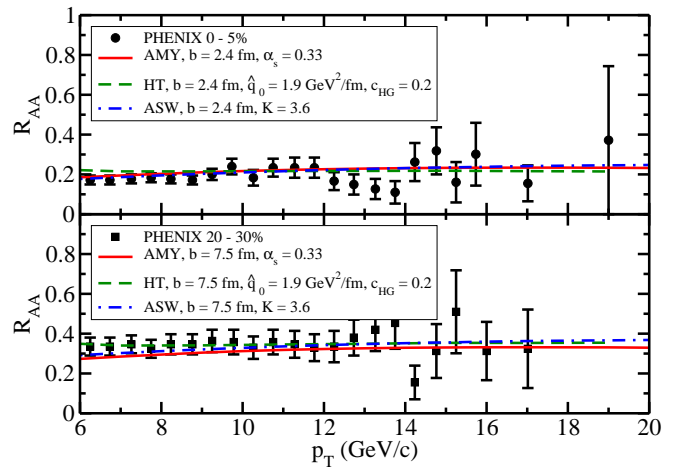}
\caption{(Color online) Nuclear modification factor $R_{AA}$ in $Au$-$Au$ collisions at 0-5\% (top) 
and 20-30\% (bottom) centrality calculated in the ASW, HT and AMY approaches compared to data from PHENIX~\cite{Shimomura:2005en}.}
\label{fig4}
\end{figure}

Figure~\ref{fig4} shows the nuclear modification factor $R_{AA}$ as a function of $p_T$ in $Au$-$Au$
collisions at 0-5\% (top)  and 20-30\% (bottom) centrality calculated in the ASW, HT and AMY 
approaches compared to data from PHENIX~\cite{Shimomura:2005en}. The parameters for the
respective calculations are fixed to one data point in the 0-5\% centrality calculation
-- the dependence on $p_T$ and centrality of the nuclear collision are then predictions
by the respective energy-loss calculations.
As can be seen, the parameters
for all three approaches (initial maximal value for the transport coefficient $\hat{q}_0$
or coupling
constant $\alpha_s$ in the AMY case) can be adjusted such that the approaches are able to 
describe the centrality dependence of the nuclear modification factor reasonably well. 
For a gluon jet, the values are $\hat{q}_0 \approx 4.3$~GeV$^2$/fm for the HT approach, $\hat{q}_0
\approx 18.5$~GeV$^2$/fm for the ASW formalism and $\alpha_s \approx 0.33$ for the AMY approach, which can be converted
into a value of $\hat{q}_0 \approx 4.1$~GeV$^2$/fm. While values of $\hat{q}_0$, have been presented up to the first 
decimal point, one should note that the error involved is never less than the experimental error (See Sec.~V for further discussion on this 
issue). Beyond this, there remain the usual uncertainties related to using a leading order hard scattering cross section, e.g., the 
choice of the appropriate scale for the structure and fragmentation functions. There also exist additional sources of error in the 
estimations of $\hat{q}$ which arise from the set of approximations used in each of the formalisms to reduce the functional 
dependence on the properties of the medium down to one parameter.

The reader will note a somewhat smaller value of 
$\hat{q}_0$ quoted for the HT formalism in Fig.~\ref{fig4}. Since, the HT approach was originally developed for DIS on a large 
nucleus, it has become customary to quote the value of $\hat{q}_0$ for a quark which is always the produced hard parton in DIS 
(see Refs.~\cite{Majumder:2007hx,Majumder:2007ne}). Besides this difference, there remain various caveats associated 
with this value of $\hat{q}$ which have been discussed in Sec.~III [in particular see the discussion surrounding Eq.~\eqref{qhat}].

For the case of the ASW formalism, we have used the relationship~\cite{Baier:2002tc},
\begin{equation}
\hat{q}_0 = 2 K \epsilon_0^{3/4},
\end{equation}
to convert the parameter $K$ in the ASW approach to $\hat{q}_0$.
In a previous estimate using this formalism~\cite{Renk:2006pk}, the value of $\hat{q}_0$ was quoted to be somewhat lower.
This is simply due to the earlier time $\tau_0 = 0.6$ fm/c at which $\hat{q}_0$ is being quoted in the current manuscript. 
In Ref.~\cite{Renk:2006pk}, $\tau_0$ was set to $1$ fm/c.

In AMY, the relationship between
$\hat{q}$ and the coupling $\alpha_s$ reads 
\begin{eqnarray}
\hat{q} = \frac{C_A g^2 T m_D^2}{2\pi} \ln\frac{q_{\bot}^{\rm max}}{m_D} \label{AMY_qhat_formula}
\end{eqnarray} 
where $q_{\bot}^{\rm max}$ is the largest transverse momentum relevant for the collinear emission.  
One estimate is that $\left( q_{\bot}^{\rm max}\right)^{2} \approx E T$, where $E$ is the energy of 
the jet, and $T$ the temperature. Evaluating  the above expression for 3 quark
flavors, $\alpha_{s} = 0.33$, a temperature of 0.4 GeV and a jet energy of 20 GeV, one obtains 
$\hat{q} =$ 4.1 GeV$^{2}$/fm. Even though this formulation is only logarithmic in the jet energy, 
it is however more suggestive than precise \cite{CHM}.
Note that the ASW value for $\hat{q}_0$ at $\tau=0.6$~fm/c and $\epsilon_0=55$~GeV/fm$^3$
lies a factor of 3.6 higher
than the Baier estimate for an ideal QGP, $\hat{q} \approx 2 \cdot \epsilon^{3/4}$
\cite{Baier:2002tc}, while the AMY estimate is in line with that from Baier, and the HT 
calculation lies about a factor of 1.6 below that  value.

The large difference in $\hat{q}_0$ values between HT, AMY and ASW has been pointed 
out previously. However, our calculation shows for the first time that this difference is not due to
a different treatment of the medium or initial state. Note that the
numbers quoted here reflect the different medium scaling laws referred to as being
the natural choices for the respective approaches, namely temperature scaling for AMY,
energy-density scaling for ASW and entropy-density scaling for HT, as discussed
in the previous section. If we choose to
perform the jet energy-loss calculation with temperature $\sim T^3$ scaling for all
three approaches, we find values for $\hat{q}_0$ being 10 GeV$^2$/fm for ASW, 2.3 GeV$^2$/fm
for HT and 4.1 GeV$^2$/fm for AMY. Likewise, if we employ energy-density scaling
$\sim \epsilon^{3/4}$, we find $\hat{q}_0 = 18.5$~GeV$^2$/fm for ASW and
$\hat{q}_0 = 4.5$~GeV$^2$/fm for HT (the AMY calculation can only be performed utilizing
temperature scaling). Both ASW and HT consistently show a rise of a factor of two in $\hat{q}_0$
when switching from temperature scaling to energy-density scaling. The different values for $\hat{q}_0$ in the
different schemes with different choices of scaling with $T$, $s$ and $\ep^{3/4}$ are presented in Table.~\ref{qhat_table}.

\begin{table}
{\color{black}
\begin{tabular}{| c | c | c | c |} \hline
\,\,$\hat{q}(\vec{r}, \tau)$ \,\, & \,\, ASW \,\,   & \,\, HT \,\, & \,\,AMY \,\,  \\
\,\,scales as \,\,&\,\, $\hat{q}_0 $  \,\,&\,\, $\hat{q}_0 $  \,\,& \,\,$\hat{q}_0 $ \,\, \\
\hline \hline 
\,\,$T(\vec{r}, \tau)$\,\, &  10 GeV$^2$/fm & 2.3 GeV$^2$/fm & 4.1 GeV$^2$/fm\\
\hline
\,\,$\epsilon^{3/4}(\vec{r}, \tau)$ \,\,& 18.5 GeV$^2$/fm & 4.5 GeV$^2$/fm &  \\
\hline
\,\,$s(\vec{r}, \tau)$ \,\,& & 4.3 GeV$^2$/fm &\,\,\,\, \\
\hline
\end{tabular}}
\caption{Values of $\hat{q}_0$, i.e., the $\hat{q}$ at $\tau = \tau_0 = 0.6$ fm/c in the cell at $\vec{r} = 0$ of the $0-5$\% centrality event, in the 
different energy loss schemes. Also presented is the variation of $\hat{q}_0$ with different choices of scaling of $\hat{q} (\vec{r},\tau)$ 
with different local intensive properties of the medium; where $T(\vec{r},\tau)$ is the temperature, 
$\epsilon (\vec{r},\tau)$ is the energy density and $s(\vec{r},\tau)$ is the entropy density at location $(\vec{r},\tau)$ in the RFD simulation. 
Given the model of the medium in AMY, $\hat{q}$ may only be calculated as a function of $T$ (see Eq.~\ref{AMY_qhat_formula}), hence the 
entries corresponding to $\epsilon$ and $s$ scaling are left blank. Calculations in the ASW scheme with $\hat{q}$ scaled with $s$ have not yet 
been performed and so the entry for  $s$ scaling has been left blank.}
\label{qhat_table}
\end{table}

\begin{figure}[tbhp]
\includegraphics[width=\columnwidth]{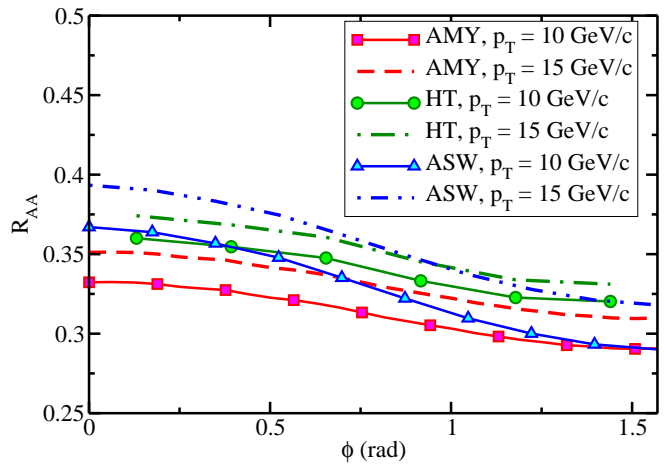}
\caption{(Color online) $R_{AA}$ as a function of azimuthal angle at $p_T=10$~GeV/c
and $p_T=15$~GeV/c for all three approaches in the 20-30\% centrality bin.}
\label{fig5}
\end{figure}

\begin{figure}[tbhp]
\includegraphics[width=\columnwidth]{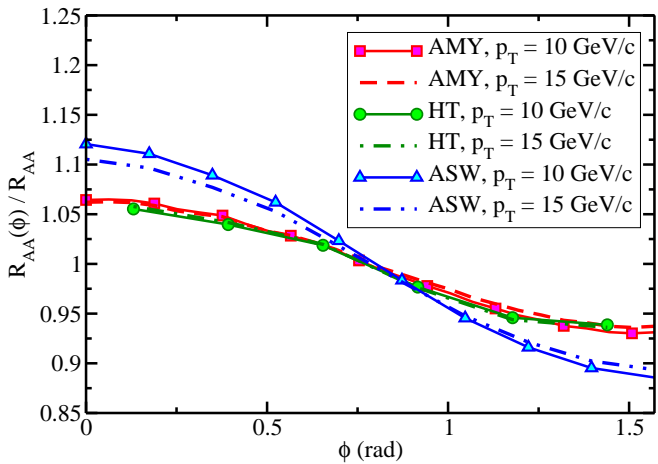}
\caption{(Color online) $R_{AA}$ as a function of azimuthal angle at $p_T=10$~GeV/c
and $p_T=15$~GeV/c  for all three approaches in the 20-30\% centrality bin,
normalized by the azimuthally averaged value of $R_{AA}$ for the respective calculations}
\label{fig6}
\end{figure}

We find that slight differences appear between the approaches when $R_{AA}$ is 
studied as a function of azimuthal angle. This can be seen in Fig.~\ref{fig5},
where $R_{AA}$ is plotted as a function of azimuthal angle at $p_T=10$~GeV/c 
and $p_T=15$~GeV/c for all three approaches in the 20-30\% centrality bin.
Figure~\ref{fig6}, shows the same calculation, but with all curves normalized by their
respective azimuthally averaged $R_{AA}$ -- we observe that for the $p_T$ bins chosen,
the AMY and HT calculations exhibit the same peak-to-valley ratio and shape, whereas
the ASW calculation shows a more pronounced difference between in-plane and out-of-plane
emission. The azimuthal spread is insensitive to variation of the transverse momentum,
which is manifest in the comparison between the solid ($p_T=10$~GeV/c) and the dashed
($p_T=15$~GeV/c) lines.

\begin{figure}[tbhp]
\includegraphics[width=\columnwidth]{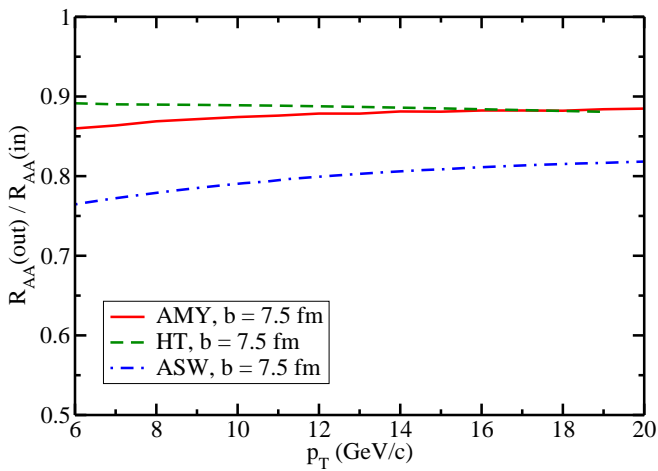}
\caption{(Color online) Ratio $R_{AA}$ for out of plane vs. in plane emission as a function of $p_T$ 
at b=7.5 fm impact parameter for all three approaches.}
\label{fig7}
\end{figure}

In order to further quantify the difference between the three approaches we calculate the 
ratio of the out of plane $R_{AA}$ over the in plane $R_{AA}$ as a function of transverse
momentum -- this is shown in Fig.~\ref{fig7}. We find that AMY and HT
exhibit the same peak to valley ratio throughout the entire range of transverse momenta, 
even though the absolute values for $R_{AA}$ differ
by approximately 10\%. The ASW calculation systematically shows a stronger azimuthal dependence
than the HT and AMY calculations - the cause of which will require a more detailed analysis
to determine.

\begin{figure}[tbhp]
\includegraphics[width=\columnwidth]{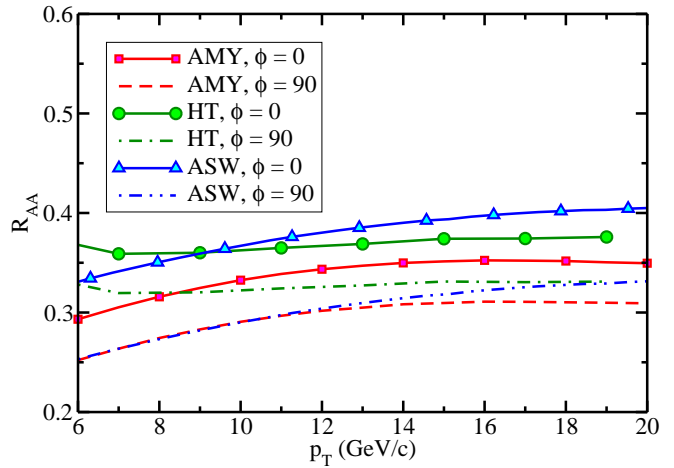}
\caption{(Color online) $R_{AA}$ for out of plane vs. in plane emission as a function of $p_T$ 
at b=7.5 fm impact parameter for all three approaches.}
\label{fig8}
\end{figure}

Note, however, that the agreement in the peak-to-valley ratio for AMY and HT does not
translate into these approaches being identical in terms of the in-plane and out-of-plane
$R_{AA}$ values vs. $p_T$: Fig.~\ref{fig8} shows that systematic 
differences on the order of 15\% exist
between all three approaches in the absolute value of $R_{AA}$ at fixed azimuthal 
angle as a function of $p_T$. 

\begin{figure}[tbhp]
\includegraphics[width=\columnwidth]{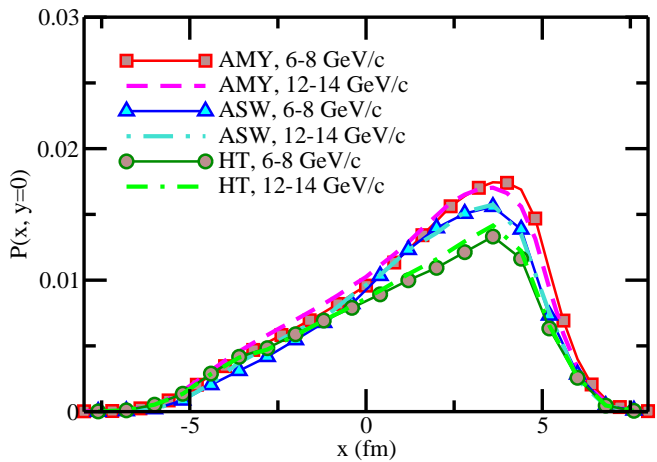}
\caption{(Color online) Survival probability $P(x,y)$  
of a hadron with 6-8 GeV/c or 12-14 GeV/c transverse momentum moving along 
the positive $x$-axis at through the center of the medium ($y=0$) in the transverse
plane as a function of $x$.}
\label{fig9}
\end{figure}

In order to investigate the spatial response of the jet energy-loss schemes to the medium, we define
the following quantity,
\begin{equation}
P(x,y) =  \frac{ T_{AB}(x,y) \cdot R_{AA}(x,y) } { \int  dx dy  T_{AB}(x,y) \cdot R_{AA}(x,y)},
\end{equation}
where the local position-dependent nuclear suppression factor $R_{AA}(x,y)$ is weighted with the nuclear overlap
probability function $T_{AB}(x,y)$. Figure~\ref{fig9}, shows $P(x,y=0)$ as a function of $x$ for a quenched jet moving
in the positive $x$ direction through the center of the medium ($y=0$).

\begin{figure}[tbhp]
\includegraphics[width=\columnwidth]{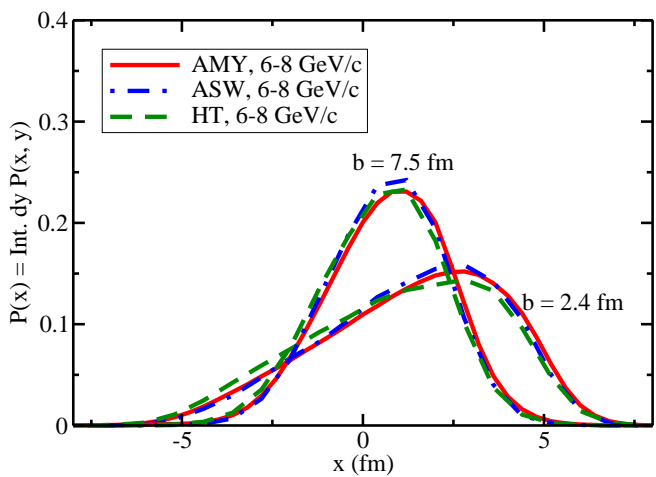}
\caption{(Color online) Escape probability of a hadron
with 6-8 GeV/c transverse momentum moving along the positive $x$-axis in the transverse
plane as a function of $x$.}
\label{fig10}
\end{figure}

Integrating the quantity $P(x,y)$ over $y$ yields the escape probability of a hadron with a transverse momentum
between 6 and 8 GeV/c originating from a quenched jet moving in the positive $x$ direction in the transverse plane as
a function of of its production vertex along the $x$-axis:
\begin{equation}
P(x)= \int {\rm dy} P(x,y)
\end{equation}
The result is shown in Fig.~\ref{fig10} --
it is remarkable how well the three different approaches agree with each other in this
quantity. Since the same hard scattering probability was used as input in all three
cases, the agreement
in $P(x)$ really shows that all three approaches yield the same suppression factor 
as a function of production vertex of the hard probe, i.e. that they probe the 
density of the medium in the same way.

\section{Normalization and further comparison to data}


As we have seen in the previous section, 
there do exist noticeable differences in the $R_{AA}$ as a function of the azimuthal
angle between the three approaches. A comparison to 
experimental data for this particular observable would thus constitute an important experimental 
input and possibly serve as a discriminator. Recently, data for $R_{AA}$ versus the 
reaction plane have become available in the $p_T=5-8$~GeV 
region~\cite{Adler:2006bw}. Unfortunately this $p_T$ range, which in terms of the data
will be dominated by the lower $p_T$ boundary, still sits in the region in
which particle production is significantly influenced by parton recombination as hadronization
mechanism~\cite{Fries:2003vb,Fries:2003kq}.
Since we regard $p_T = 6$ GeV  as the lower limit of the applicability of jet quenching calculations, 
a  comparison may not be completely out of place, but would carry large uncertainties with it.

However, the data from run-2 of the PHENIX collaboration~\cite{Adler:2006bw} which was used to deduce the $R_{AA}$ versus the 
reaction plane demonstrates an integrated $R_{AA}$~of~$0.41~\pm~0.03(stat)~\pm~0.06(sys)$ in the 20-30\% centrality events, in noticeable contrast to 
the value of $0.35~\pm\sim0.04(stat)~\pm\sim0.03(sys)$ as seen in Fig.~\ref{fig4} from the run-4 data set. 
While the two data sets agree within systematic errors, the discrepancy between the two
is too large for a meaningful comparison of our calculation, which was fit 
to the 
run-4 data set.

\begin{figure}[tbhp]
\includegraphics[width=\columnwidth]{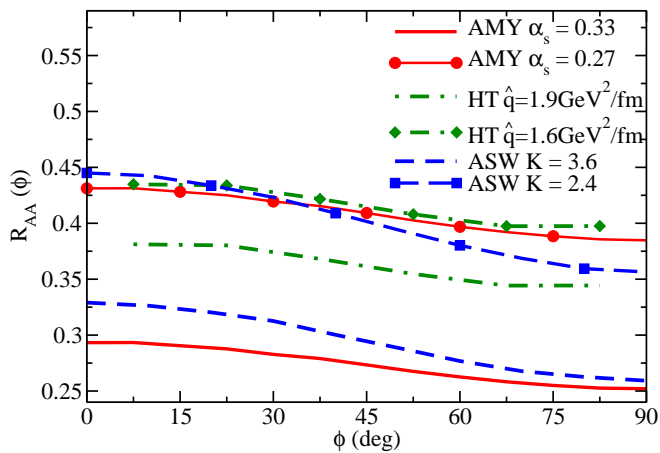}
\caption{(Color online) $R_{AA}$ vs. reaction plane in the 20-30\% centrality event at $p_T = 6$ GeV 
for different choices of the single fit parameters $\hat{q}, K, \alpha_s$}
\label{fig11}
\end{figure}

An estimate of the variation of the fit parameters required to encompass both data sets leads to differences 
of the order of $20 - 40$\% in $\hat{q}$. Plotted in Fig.~\ref{fig11}, are the predictions for the $R_{AA}$ versus reaction 
plane for the standard values of the fit parameters obtained from the comparison with the run-4 data set in 
Fig.~\ref{fig4}. Also plotted are readjusted plots for the $R_{AA}$ versus the reaction plane where the 
single fit parameters in each of the models was tuned such that the integrated $R_{AA}$ in the 20-30\%
centrality bin achieved a value of 0.41. The new values of the fit parameters (included in the figure) are 
$\hat{q} = 1.6$ GeV$^2/$fm for the HT, $K=2.4$ for the ASW and $\A_s = 0.27$ for the AMY calculations  respectively.  
One should note that a simple renormalization of our $R_{AA}$
vs. $\phi$ curves
to the data would not be appropriate, since the value of $\hat{q}$ 
affects the magnitude of the azimuthal spread.

A detailed and meaningful theory-experiment comparison, 
encompassing different data sets as well as their
respective statistical and systematic errors in a proper fashion will require a sophisticated
statistical analysis beyond the scope and aim of the work presented here. Such an analysis
has been demonstrated for one particular theory calculation of
inclusive $R_{AA}$ vs. $p_T$ compared to one experimental data
set in \cite{Adare:2008cg}. The feasibility of extending such an analysis to multiple 
data sets, observables and theory calculations has yet to be determined. 

\section{Summary and Conclusion}


In summary, we have calculated the modification of hard jets in a 3D hydrodynamic
medium, in three different 
approaches which were constrained to use the same initial structure functions, the same final 
vacuum fragmentation functions, the same nuclear geometry and identical three dimensional 
evolution of the produce dense matter. In this first, unified, attempt to understand jet modification 
in dense matter, the focus was restricted to single inclusive observables. The nuclear modification 
factor [Eq.~\eqref{raa}] was computed 
as a function of the transverse momentum, centrality of collision, as well as, the angle with respect 
to the reaction plane. This was followed by a more detailed, though purely theoretical, analysis of jet origin 
distribution for the $R_{AA}$ as a function of the reaction plane, as well as, the $R_{AA}$ for jet origins 
restricted to lie on a narrow belt on the reaction plane. 

In the comparisons above, both the HT and the ASW schemes have been simplified to the point that 
all predictions depend on only one tunable parameter: this is the $\lc FF \rc$ correlator in the HT approach 
and the $K$ parameter in the ASW approach. In the most rigorous formulation of AMY, there exist no 
free parameters except for the temperature; this however, has already been specified by the RFD simulation.
In the phenomenological application of the AMY approach used here, the strong coupling constant 
is treated as a parameter: It has, thus, been disassociated from the temperature. 

These single free parameters from all three approaches were tuned to fit one data point, usually chosen 
as the integrated $R_{AA}$ at 8 GeV in the 0-5\% centrality events. The data used for this comparison as shown 
in Fig.~\ref{fig4} were taken from the PHENIX run-4 data set \cite{Shimomura:2005en}. 
Our comparison shows
that under identical conditions (i.e. same medium evolution, same choice
of parton distribution functions, scale etc.) all three jet energy-loss schemes yield 
very similar results. This finding is very encouraging since it indicates that the technical
aspects of the formalisms are well under control. However, we need to point out that there
still exists a {\em puzzle} regarding the extracted value for the transport coefficient 
$\hat{q}_0$, which spans a factor of four from a value of 2.3~GeV$^2$/fm for the HT
approach on the lower end, to 4.1~GeV$^2$/fm for AMY and  $10$~GeV$^2$/fm for 
ASW on the high end, when using the same temperature scaling law for all three
approaches. 
While the discrepancy among these approaches is not new, our work has been able to decisively 
rule out
differences in the medium evolution or initial setup as a cause for the differing values
of $\hat{q}$. We are led to conclude that these remaining differences are due, to
the different approximations applied, to the different energy scales involved, to the different assumptions on the structure of the QCD matter, inherent in these
different approaches. 

There exist multiple future directions for the systematic
and unified approach to jet modification in dense matter
presented here. 
Due to the assumption of a thermalized plasma, elastic energy loss may  be straightforwardly 
included in AMY. Including elastic energy loss, however, represents a significant extension to the 
HT and ASW approaches which has, only recently, been undertaken and 
thus, this topic has not been included in the comparisons presented in this article.
The current effort was restricted to single inclusive observables; hence, the simplest extension will be to apply a 
similar analysis to both single and multi-particle observables in tandem. Such comparisons will undoubtedly 
lead to stronger constraints on the formalism and hence deeper insights in the nature of the theory of jet modification 
used. Another direction is to use a somewhat different initial condition and equation of state for the medium evolution. A natural 
extension in this direction is to the study of jet modification in viscous fluid dynamical simulations. Viscous 
simulations, necessarily seem to require an initial state with greater spatial anisotropy. We believe, that it is 
in this direction that measurements and theoretical calculations of the $R_{AA}$ versus the reaction plane will have most relevance, 
as a means to discriminate between different initial state profiles. The approximations which have resulted in the 
reduction of formalisms such as the HT and the ASW to a dependence on only one parameter will eventually 
have to be relaxed. The different parameters in these schemes represent actual physical properties of the 
produced matter which may indeed be measurable given a detailed and extensive set of experimental measurements.

\section{Acknowledgments} 
This work was supported in part by the U.~S.~Department of Energy (grant DE-FG02-05ER41367) (S.A.B., A.M., C.N.), in part
by the Academy of Finland, Project 206024 (T.R.), and in part by the Natural Sciences and Engineering Research Council
of Canada (C.G., G.Y.Q., J.R.). C.G., G.Y.Q., and J.R. are happy to acknowledge useful discussions with S. Caron-Huot, S. Jeon and G. D. Moore.
The authors would also like to thank B.~M\"{u}ller for a careful reading of the manuscript.

\bibliographystyle{h-physrev}
\bibliography{SABrefs}

\end{document}